# Hemodynamics of a Bileaflet Mechanical Heart Valve with Different Levels of Dysfunction

**Fardin Khalili, Gamage PPT and Mansy HA\***

*Department of Mechanical and Aerospace Engineering, University of Central Florida, USA*

**\*Corresponding author:** Fardin Khalili, Biomedical Acoustics Research Laboratory (BARL), Department of Mechanical and Aerospace Engineering, College of Engineering and Computer Science, University of Central Florida, ENGR 1, Room 428, 12760 Pegasus Blvd, Orlando, FL 32816, USA, Fax (407) 823-0208; Tel: (616) 920-2661; Email: fardin@knights.ucf.edu

**Abstract**

Heart disease is one of leading causes of mortality worldwide. Healthy heart valves are key for proper heart function. When these valves dysfunction, a replacement is often necessary in severe cases. The current study presents an investigation of the pulsatile blood flow through a bileaflet mechanical heart valve (BMHV) where one leaflet is healthy and can fully open and the other leaflet cannot fully open with different levels of dysfunction. To better understand the implications that a dysfunctional leaflet has on the blood flow through these valves, analysis of flow characteristics such as velocity, pressure drop, wall shear stress and vorticity profiles was performed. Results suggested that leaflet dysfunction caused increased local velocities, separation regions and wall shear stresses. For example, the maximum velocity increased from 2.53 m/s to 4.9 m/s when dysfunction increased from 0% to 100%. The pressure drop increased (by up to 300%) with dysfunctionality. Results suggested that leaflet dysfunction also caused increased wall shear stresses on the valve frame where higher stresses developed around the hinges (at 75% and 100% dysfunctions). Analysis also showed that increased dysfunctionality of one leaflet led to higher net shear forces on both the healthy and dysfunctional leaflets (by up to 200% and 600%, respectively).

**Abbreviations:** BMHV: Bileaflet Mechanical Heart Valve; CFD: Computational Fluid Dynamics; LDA: Laser Doppler Anemometry; PIV: Particle Image Velocimetry; WSS: Wall Shear Stresses

## Introduction

Heart disease is one of the leading causes of death. Treatment of certain types of heart disease involves surgically implanting mechanical heart valves. Bileaflet mechanical heart valves (BMHVs) are commonly used for valve replacement because their design can minimize flow disturbances [1]. However, the rate of artificial valve dysfunction is 0.2-6% patient/year [2]. Analysis of blood flow characteristics such as velocity, vortex formation and turbulent stresses are of importance to identify potential blood cell damage [3,4], especially due to complex and unsteady flow in the valve hinges [5,6]. Woo et al. [7] and Hasenkam [8] showed that the high-velocity flow through the valve resulted in higher shear stress levels at the valve hinges and downstream of the valve and consequently, restricts the leaflets motion. Moreover, high velocity and shear stresses were observed in leakage jets from BMHV hinges which were associated with platelet activation and also thrombosis due to increased residence time in the hinges. These life-threatening complications may lead to dysfunction in one or both leaflets of BMHVs [9]. Hence, analysis of cardiac sounds [10,11] and flow dynamics [12-15] are active research areas, which may lead to better identification of the complications and dysfunction issues and consequently, help to improve valve design.

Previous studies analyzed the blood flow for different BMHV configurations. King et al. [16] reported that the leaflet opening angle affects the blood flow behavior and concluded that a larger leaflet opening angle reduced the pressure drops across the valve. Pulsatile Newtonian blood flow was studied for fixed leaflets at different levels of dysfunction [2]. Another study illustrated some of the characteristics of pulsatile Newtonian and non-Newtonian blood flow through moving healthy leaflets and aortic root sinuses [17].

Previous studies of BMHV have shown that the blood flow moves through three orifices and creates three jets that cause stronger shear stresses than the case of natural valves [1,16]. Furthermore, studies suggested that vortices become more prominent and wall shear stresses increase with leaflet dysfunction [2]. Increased wall shear stresses are important to detect as it can contribute thrombus formation [1]. Moreover, a leaflet dysfunction can cause backflow. Several different techniques to study blood flow in BMHV were implemented in previous studies. This included CFD (Computational Fluid Dynamics) and experimental methods such as PIV (Particle Image Velocimetry), Video Analysis, LDA (Laser Doppler Anemometry) [2,16,17]. In the current study, CFD is used to model the valves and blood flow as the software allows relatively easy and economical conduction of parametric studies. The objective of this study is to investigate the flow around BMHV at different levels of leaflet dysfunction by:

a) Documenting the relevant flow structures using streamlines and vorticity information

b) Quantifying the changes in pressure drop

c) Calculating wall shear stresses on the valve frame and shear forces on the leaflets

d) Comparing the results of Newtonian and non-Newtonian fluids.

This study may provide data that can enhance our understanding of the implications dysfunctional leaflets, which can lead to improved design of BMHV.

## Models and Methods

In this study, the computational domain was divided into four regions sequentially in the flow direction: upstream, heart valve, aortic root sinuses and downstream. The heart valve geometry (Figure 1a) investigated was chosen to be similar to previous studies [2,16-20]. Figure 1b shows the asymmetric aortic root sinuses geometry (with inlet diameter of 0.023 m), which was generated using information extracted from a previous experimental study [21] and was represented as an epitrochoid. Creating a realistic geometry of the aortic sinuses is important for appropriate internal flow field analysis [13,22]. In the current study, the top (functional) leaflet was assumed to be fixed in the fully open position while the bottom leaflet was fixed at different levels of dysfunctionalities. Since the dynamics of the leaflet opening and closure were not simulated, the data presented in the results section will focus on the fully opening period. Figure 1c shows the side cross section of the BMHV at different levels of leaflet dysfunction of 0, 25, 50, 75 and 100%. The positive direction of the shear forces was chosen to be in the flow direction from inlet to outlet. In this study, a commercial CFD software (STAR-CCM+, CD-adapco, Siemens, Germany) was utilized to perform the unsteady simulation for one complete cardiac cycle (simulation time, T = 0.866s) with a time step = 0.5 ms and 25 iterations per time step. The in vitro CFD analyses were performed for a pulsatile flow through a three-dimensional BMHV with fixed leaflets. The inlet velocity corresponding to cardiac output of 5 L/min and heart rate of 70 bpm with systolic phase duration of 0.3 s as illustrated in [19]. The peak inflow velocity was ~1.2 m/s and the density of blood was set to ρ = 1080 kg/m$^3$. The blood also was considered to be both Newtonian with the viscosity of 0.0035 kg/(m.s) and non-Newtonian based on generalized Carreau-Yasuda model [14]. This lead to an inlet peak Reynolds number $\left(Re_{peak} = \dfrac{\rho U_{peak} d_{inlet}}{\mu}\right)$ of 8516 and a $W_0$ is Womersley number, $\left(W_0 = \dfrac{d}{2}\sqrt{\omega \rho / \mu}\right)$; $\omega = \dfrac{2\Pi}{T}$ is the frequency of pulsatile flow equals to 17.21. Moreover, high quality polyhedral mesh was generated in the flow domain, especially in the heart valve and aortic sinuses regions, close to the wall and the leaflet surfaces, Figure 2 in order to maintain y+ ≪ 1 (y+ = 0.46 at the peak flow).

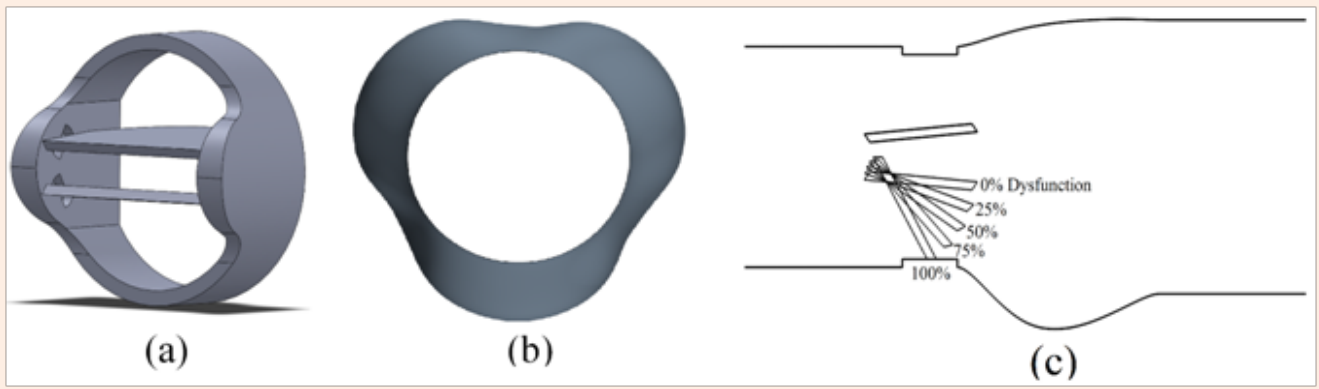

**Figure 1:**
a. Bileaflet mechanical heart valve
b. Aortic root sinuses
c. Different levels of dysfunction.

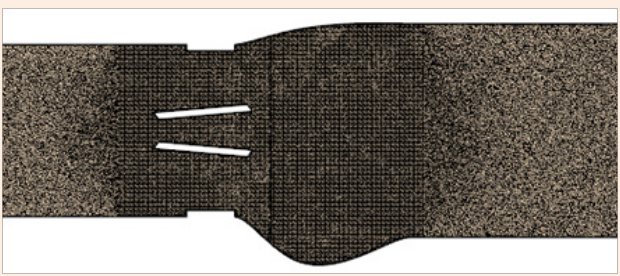

**Figure 2:** High quality mesh generated in the flow domain especially close to the leaflets and aortic sinuses.

## Validation

The normalized velocity profile along a line located 7 mm downstream of the healthy (0% dysfunction) valve is shown in Figure 3. The velocity profiles obtained in previous experimental study that considered a valve with similar geometry and flow conditions [23] are also shown in the same figure. The results from the current study are in good agreement with the previous experimental and computational data. To quantify the difference between our computational and the experimental results, the root-mean-square (RMS) of the difference in velocity was calculated. The RMS of the velocity difference was 6.58% of the maximum velocity, suggesting agreement between the results of the current study and the measured values.

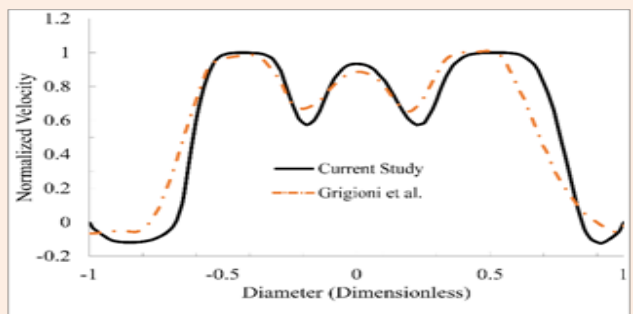

**Figure 3:** Validation of the computational methods by comparing the velocity distribution in the current study with a previous experimental study [20].

## Results and Discussion

Figures 4a2 to 4a5 shows the streamlines at the peak systolic time of 90 ms, where the color represents the velocity magnitude. For 0% dysfunction (Figure 4a1), the blood flow seemed to have a narrower velocity range (i.e. more uniform velocity) compared to the higher levels of dysfunction (Figures 4a2 to 4a5). Figure 3a1 also shows relatively smaller flow separation in the wake region downstream of the leaflets as would be expected. The flow reattachment also happened closer to the exit of the aortic sinuses. The velocity magnitude in the orifices increased, especially in the bottom orifice, as leaflet dysfunction increased. Figure 4 also suggests that an increased leaflet dysfunction may increase the potential for development of higher levels of disturbances in the flow and possibly increased turbulence. This data also suggested that more intense vortical structures start to appear in the valve and sinus regions during the acceleration phase (e.g., 60 to 90 ms). Figure 4b1 to 4b5 shows vorticity at different levels of dysfunction. Vorticity increased with dysfunctions and spread downstream of the leaflets. Conversely, lower levels of vorticity occurred in the sinus downstream of the dysfunction leaflet at 100% dysfunctions, which can be because the obstruction caused by the dysfunction created a low velocity region behind that leaflet. While Figure 4 shows information for t= 90ms, flow structures were also examined for all times between 60 to 250 ms and were found similar to those shown in Figure 4.

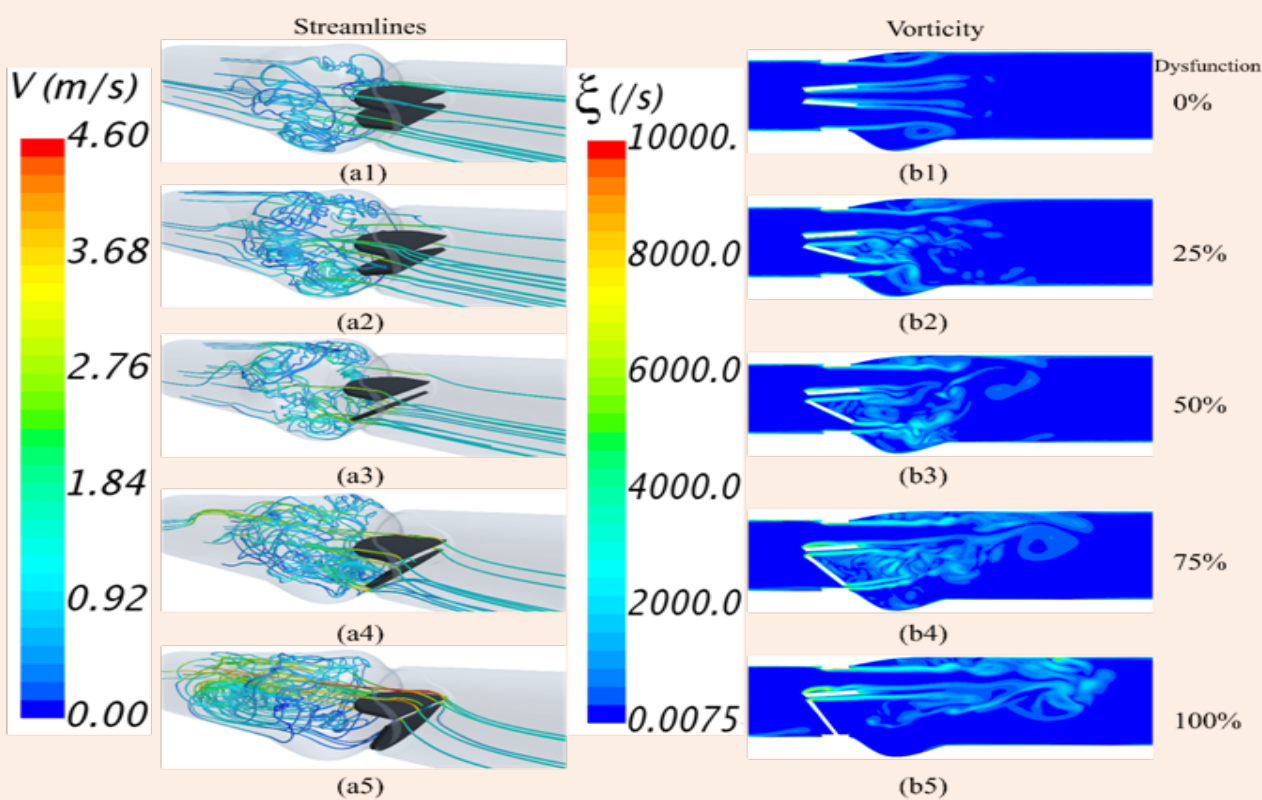

**Figure 4:** Flow field around the valve

    a. Streamlines

    b. Vorticity

The maximum velocity in the flow domain was determined and was found to occur at the time of the peak systolic velocity. Results suggested that the maximum flow velocity increased with dysfunction. For instance, as dysfunctionality increased from 0% to 100%, the maximum velocity increased from 1.63 to 3.3 m/s and from 1.52 to 3.65 m/s in the middle and top orifices, respectively. In addition, the maximum flow velocity increased from 2.52m/s to 4.89m/s. Higher velocities and flow separation at the leaflet surfaces were accompanied by growing eddies and vorticity downstream of the valve (Figure 4). Figure 5 shows the pressure drop (during the cardiac cycle) for different levels of dysfunction. The figure shows that the pressure drop tended to increase with dysfunctionality. The maximum pressure drop (around t=0.06 s in Figure 5) increased from 20mmHg to 63mmHg (~2300 Pa to ~8500Pa). Note that since leaflets are fixed in the current study, the pressure drop in the fully open period is most relevant.

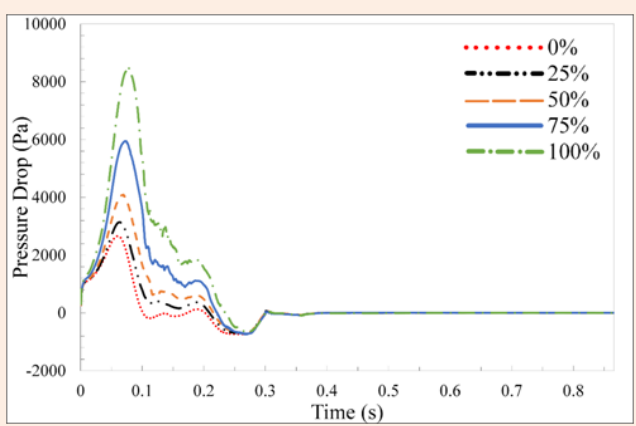

**Figure 5:** Pressure drop from inlet to outlet at different levels of dysfunction.

Figure 6 shows the wall shear stresses (WSS) on the valve frame at different levels of dysfunction for Newtonian (Figure 6a1 to a5) and non-Newtonian (Figure 6b1 to b5) flow conditions. Wall shear stresses increased with dysfunction, which was accompanied by increased velocity in the orifices. The case of at 0% dysfunction (Figure 6a1) was associated with lower WSS on the valve frame. At 50% and 75% dysfunctions, wall shear stresses increased on the valve wall downstream of the bottom orifice where flow with higher velocities passed through the orifice. At 100% dysfunction, lowest WSS was observed on the surface at the bottom valve surface. In addition, higher WSS developed around the hinges and frontal surface of the valve with dysfunctions, especially at 75% and 100% dysfunctions. Identification of areas of high WSS is important as it is associated with increased risk of thrombus formation [1]. As shown in Figure 6, WSS magnitudes for Newtonian flow were similar to those for non-Newtonian flow. The maximum difference between the two cases was less than 2%. For example, the maximum WSS at 100% dysfunction was ~951 Pa and ~954 Pa for Newtonian and non-Newtonian flows, respectively. Therefore, we can conclude that the Newtonian flow assumption is appropriate for calculating WSS on valve frame.

Table 1 illustrates the maximum values of the averaged wall shear stress applied on the heart valve frame which occurred at the peak systole. The averaged and maximum wall shear stresses on the valve frame at the peak systole increased with dysfunction. The information regarding the location of the highest WSSs in the flow domain is presented in this table. Helicity is proportional to the flow velocity and vorticity and indicates the potential for development of helical flow. The data indicate that the helicity increased with dysfunction and peaked around peak systolic velocity time. Shear forces on the top and bottom leaflets were also studied for Newtonian and non-Newtonian flows. Figure 7 shows the shear forces on the top and bottom leaflets at 50% dysfunction. The results indicate that the shear forces on top and bottom leaflets for Newtonian flow is similar to non-Newtonian

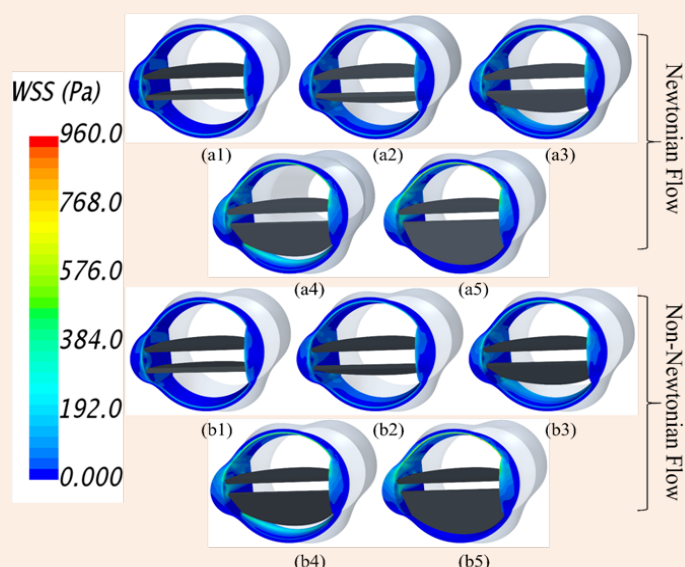

**Figure 6:** Wall shear stresses on valve frame

  a) Newtonian flow

  b) Non-Newtonian flow (1: 0%; 2: 25%; 3: 50%; 4:75% and 5: 100% dysfunction)

flow. The maximum difference between the two flow conditions was less than 1%. While Figure 7 shows shear forces for 50% dysfunction, similar trends were seen at different levels of dysfunction.

**Table 1:** Averaged and maximum wall shear stresses (WSSs) on the valve frame, location of the highest WSSs in the flow domain and the maximum helicity in the aortic sinuses.

| Dysfunction (%) | Averaged WSS at Peak Systole(Pa) | Maximum WSS at Peak Systole(Pa) | Location of the Highest WSSs | Maximum Helicity in Aortic Sinuses (m/s^2) |
|---|---|---|---|---|
| 0 | 27.86 | 241.67 | Leading edge of the leaflets | 2088.01 |
| 25 | 38.71 | 326.77 | Leading edge of the leaflets | 5209.04 |
| 50 | 48.08 | 448.96 | Leading edge of the leaflets, trailing edge of the dysfunctional leaflet and top leaflet hinges | 5936.48 |
| 75 | 65.34 | 666.61 | Inner surface of the Valve frame close to the trailing edge of the dysfunctional leaflet, bottom surface of the top leaflet and top leaflet hinges | 8328.39 |
| 100 | 50.24 | 952.78 | Upper half of the valve frame, top leaflet top and bottom surfaces and top leaflet hinges | 9794.89 |

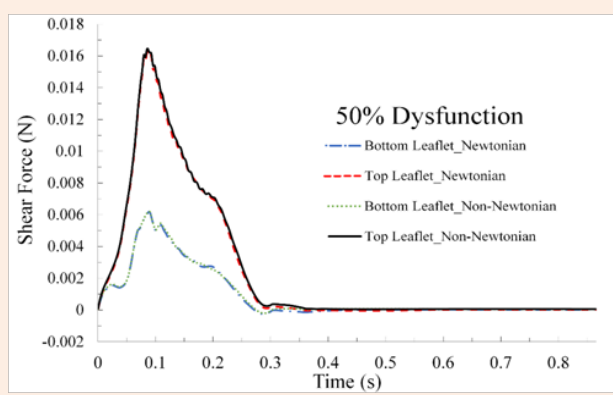

**Figure 7:** Shear forces on top and bottom leaflet at 50% dysfunction for Newtonian and non-Newtonian flows.

## Conclusion

In this study, blood flow through a bileaflet mechanical heart valve was analyzed at different levels of leaflet dysfunction. Results suggested increased vortical structures and velocities with dysfunction. For instance, as dysfunctionality increased, the maximum velocity increased from 1.63 to 3.3 m/s and from 1.52 to 3.65 m/s in the middle and top orifices, respectively. Higher velocities and flow separation at the leaflet surfaces were accompanied by growing eddies and vorticity downstream of the valve. The pressure drop was found to increase with dysfunctionality (from 20 mmHg to 63 mmHg). Results suggested that leaflet dysfunction caused increased wall shear stresses on the valve frame. Higher stresses developed around the hinges (at 75% and 100% dysfunctions) and some part of the valve surfaces at the bottom orifice (at 25% and 50% dysfunctions). Identification of regions of elevated shear stress is important since this can increase thrombus risk. Analysis also showed that increased dysfunctionality of one leaflet led to higher net shear forces on both the healthy and dysfunctional leaflet (by up to 200% and 600%, respectively). These results were similar for both Newtonian and non-Newtonian flow suggesting that the assumption of Newtonian flow, which can reduce computational cost, would be valid.